\documentclass[aps,pre,a4paper,reprint]{revtex4-1}
\usepackage[dvips]{graphicx}
\usepackage[dvipsnames]{xcolor}
\usepackage{soul}
\usepackage{epsfig}
\usepackage{dcolumn}
\usepackage{bm}
\usepackage{amsmath}
\usepackage{amssymb} 
\usepackage{hyperref}
\usepackage{natbib}
\usepackage[normalem]{ulem}

\begin{document}

\title{Power law relationship between  diffusion coefficients in  multi-component glass forming liquids}

\author{
Anshul D. S. Parmar$^{1,2}$, Shiladitya Sengupta$^{3}$, Srikanth
Sastry$^{1}$
}

\affiliation{
$^{1}$ Theoretical Sciences Unit, Jawaharlal Nehru Centre
for Advanced Scientific Research, Bengaluru, India.\\
$^{2}$ Tata Institute of Fundamental Research, 36/P, Gopanpally Village, 
Serilingampally Mandal,Ranga Reddy District, Hyderabad, 500107, India. \\
$^{3}$ Dept. of Fundamental Engineering, Institute of Industrial
Science, The University of Tokyo, Komaba 4-6-1, Meguro-ku, Tokyo
153-8505, Japan.
}


\begin{abstract} {The slow down of dynamics in glass forming liquids as the glass transition is approached has been characterised through the Adam-Gibbs relation, which relates relaxation time scales to the configurational entropy. The Adam-Gibbs relation cannot apply simultaneously to all relaxation times scales unless they are coupled, and exhibit closely related temperature dependences. The breakdown of the Stokes-Einstein relation presents an interesting situation to the contrary, and in analysing it, it has recently been shown that the Adam-Gibbs relation applies to diffusion coefficients rather than to viscosity or structural relaxation times related to the decay of density fluctuations. However, for multi-component liquids -- the typical cases considered in computer simulations, metallic glass formers, {\it etc.} -- such a statement raises the question of which diffusion coefficient is described by the Adam-Gibbs relation. All diffusion coefficients can be consistently described by the Adam-Gibbs relation if they bear a power law relationship with each other. Remarkably, we find that for a wide range of glass formers, and for a wide range of temperatures spanning the normal and the slow relaxation regimes, such a relationship holds. We briefly discuss possible rationalisations of the observed behaviour. }
\end{abstract}


\maketitle

\section{Introduction}
\label{sec:intro}
An understanding of the dramatic slow down of relaxation dynamics upon approaching the glass transition is central to understanding the behaviour of glass forming liquids. While most theoretical descriptions focus on slow down that is common to various quantities that may be used to describe it, it is well appreciated that the variety of characteristic time scales that one may study, their heterogeneity, {\it etc}, display rich and complex variations \cite{berthier2011theoretical,berthier2011dynamical,starr2013relationship,greer2013fragility,karmakar2014growing,karmakar2015length}. 
Some commonly studied measures of relaxation are the diffusion coefficient ($D$), the
viscosity ($\eta$), and the $\alpha$-relaxation time ($\tau_\alpha$)
measured from density correlation functions at a given probe
length scale - {\it e.g.} the self-intermediate scattering function
$F_s(k,t)$ at a wave number $k$ corresponding to the peak of the
static structure factor. At normal liquid temperatures, the temperature dependence of these quantities are expected to be related, {\it e. g.} through the application of the Stokes-Einstein relation. 
The  Stokes-Einstein (SE) relation between the
translational diffusion coefficient of a diffusing Brownian particle of mass
$m$ and radius $R$ and the shear viscosity of the liquid medium in
which it is diffusing:
\begin{equation}
  D = {m k_B T \over c \pi R \eta}
\end{equation}\label{eqn:SE}
where $c$
is a constant determined by the boundary condition (stick or slip) at the surface of the Brownian particle, and $T$ is the
temperature of the liquid. This relation is applied to the self
diffusion coefficients of liquids and is found to be obeyed quite well
\footnote{The Stokes' law for the drag force was originally derived
  for \emph{Brownian tracer} particles much larger than the particles of the
  surrounding medium, which was modelled as a continuum fluid obeying
  macroscopic hydrodynamics. Thus it is quite surprising that this
  relation actually holds at all for \emph{self}-diffusion coefficient of the
  fluid particles, even at high temperatures.} 
However, a hallmark of the glassy dynamics of a liquid approaching the glass transition is the decoupling of
different time scales, exemplified by the breakdown of the
Stokes-Einstein relation (SEB) \cite{RosslerGrp1990,FujaraEtal1992,HodgdonStillinger1994,TarjusKivelson1995,EdigerGrp1996,DOUGLAS1998,sillescu1999,XiaWolynes2001,BonnKegel2003,berthier2004time,Jungetal2004,Chong2008,ChongKob2009,EdigerGrp2011,Langer2011,CharbonneauEtal2013,SenguptaEtal2014,BhowmikEtal2016}. Below a characteristic temperature, the relationship between the diffusion coefficient and the viscosity above does not hold, and instead, it is observed that the diffusion coefficient is related to the viscosity through a fractional power ( {\it i. e.}, $D \propto \eta^{-\xi}$, with $\xi < 1$). A similar statement applies to the relationship between the diffusion coefficient and relaxation time scales $\tau(k)$ obtained at different wave vectors  $k$, below breakdown temperatures that decrease with a decrease in $k$ \cite{ParmarEtal2017AG}. 

%
%

The slow down of dynamics is rationalized in terms of a decrease in the configurational entropy  ($S_c$) by the Adam-Gibbs relation \cite{AG1965}, which may be written in the form 

\begin{equation}
  X = X^0 \exp \left({A_X \over TS_c}\right)
\end{equation}\label{eqn:AG}
where $X$ is a characteristic relaxation timescale  and $A_X / S_c$ is the corresponding
activation energy at a low temperature $T$. With the configurational entropy saturating to a high temperature value, 
such a form may also further be seen to provide an Arrhenius form at high temperatures, and the Adam-Gibbs parameters may be 
 related to the activation energy in the Arrhenius regime at high temperature and to the fragility of the glass formers \cite{Sastry2001, SenguptaEtal2011, Parmar2015,
  ParmarEtal2017}. The Adam-Gibbs relation has been extensively analysed using both experimental and simulation data (see \cite{Angell1997a,Sastry2001,Scala2000b,Karmakar2009d,Sengupta2013e,Banerjee2014b} for some examples). In addition to analysing the case of "fragile" liquids which exhibit strongly non-Arrhenius temperature dependence of dynamics, the Adam-Gibbs relation has also been employed to rationalise more complex behaviour of substances like water, which exhibit a fragile-to-strong crossover \cite{Ito1999d,Sastry1999,Starr2000c,Starr2003a}. The ability of the Adam-Gibbs relation to capture dynamical changes in systems with different spatial dimensions has also been tested \cite{sengupta2012adam}. It is found that $3$ and $4$ dimensional systems obey the Adam-Gibbs relation well, but $2$ dimensional systems exhibit marked deviations, which bear further scrutiny. 
  
The relationship between the configurational
entropy and the activation energy for structural relaxation and hence
the dynamics is a central idea to various more detailed entropy-based
theories of glass transition such as the Random First Order Transition
(RFOT) theory \cite{RFOT1989, RFOT2007, RFOT2014} and the Generalized
Entropy Theory \cite{GET2008,Freed2014a}. Nevertheless, whether this relation should apply primarily to diffusion or structural relaxation as captured by viscosity or the $\alpha$-relaxation time, has not been satisfactorily addressed. This question becomes relevant when the different dynamical quantities decouple, which does indeed occur in glass forming liquids. However, in view of the observed fractional  Stokes-Einstein relation, it has been argued that the diffusion coefficient and viscosity, for example, may obey the Adam-Gibbs relation consistently, with different activation energies $A_X$ \cite{SenguptaEtal2013SEB} -- the fractional exponent simply scales the activation energy in the activated Adam-Gibbs expression. However, such a rationalisation is applicable when one limits attention only to the temperature regime where a single  (fractional or inverse) power law relationship holds. If one considers a broad enough temperature range which encompasses the regime in which the diffusion coefficient and viscosity either bear the relationship predicted by the Stokes-Einstein relation, and a regime in which one has the breakdown of the Stokes-Einstein relation, clearly, the Adam-Gibbs relation cannot apply to both diffusion and viscosity simultaneously. Considering such a situation, it was recently shown that indeed, the Adam-Gibbs relation applies to diffusion over the broadest range of temperatures, whereas it is violated for viscosity and relaxation times defined through the self intermediate scattering function, in the presence of the breakdown of the Stokes-Einstein relation \cite{ParmarEtal2017}. The statement that the Adam-Gibbs relation applies primarily to diffusion needs theoretical understanding, but it immediately raises another question, which we address in this manuscript. 

Many glass-formers are multi-component substances, with self diffusion coefficients defined for each component. In such a case, the question 
of which diffusion coefficient(s) the Adam-Gibbs relation applies to arises naturally. This would be a non-issue if the diffusion coefficients bear a constant proportionality, such as what one may expect for Brownian particles from the Stokes-Einstein relation, but it is observed that this is not the case, and the ratio of diffusion coefficients is temperature dependent \cite{schober2016heterogeneous}. Barring constant proportionality, the only other scenario in which the different self diffusion coefficients may consistently obey the Adam-Gibbs relation is that they bear a power law relationship with each other. To test whether such is the case, we have examined self diffusion coefficients in a large range of glass formers, with data obtained both from simulations and experimental measurements. In each case, we have examined data for a wide range of temperatures, from above the `onset' temperature\cite{sastry1998c,sastry2000l,Banerjee2017d} of slow dynamics to well below it. Remarkably, we find that in such a wide range of temperatures, a power law relationship is found to hold between the self diffusion coefficients of component species. The fractional exponent is not universal, but varies from values as low as $0.52$ to values close to $1$. The details of the models and systems we consider are provided in the next section, followed by a discussion.

The rest of the paper is organized as follows: In section \ref{sec:models} we
present the data as well as summarize the details of all the
glass-forming systems surveyed here. In section
\ref{sec:disc} we discuss the implications of our findings. 

\begin{table*}[htp]
  \begin{center}
        \caption{Power law behavior of diffusivity of different components for various glass forming liquids. ``Dim'' denotes the spatial
      dimension, and ``R'' denotes the radius of a species. Indices $s$ and $f$ denote slow and fast components, respectively. Diffusion coefficient $D^{0}$ corresponds to high-temperature diffusivity. More details about the glass-formers are provided in Sec. \ref{sec:models}. }
    \label{tab:table1}
    \begin{tabular}{c|c|c|c|c|c|c}
      \textbf{Acronym}& \textbf{Description} & \textbf{Dim} & \textbf{Exponent}
      & $R_{slow} : R_{fast}$ & $D_{slow}^0$ & $D_{fast}^0$ \\
      & &  & ($\gamma$) & & &\\
      \hline
      3DZrCu \cite{kluge2004diffusion}& Embedded atom model for $Zr_{67}Cu_{33}$ & 3 & 0.66 & 206:145 & 2.45E-9 &3.57E-9 \\
      3DZrNi \cite{teichler2001structural}& 50:50 binary mixture of  $Zr_{0.5}Ni_{0.5}$ & 3 & 0.77 & 206:149  & 1.82E-11 & 4.85E-11 \\
      3DSS & 50:50 mixture of soft spheres & 3 & 0.78 &	1.4:1 & 1.02E-3 & 1.41E-3\\
      3DLJpoly \cite{murarka2003diffusion}& Polydispersed LJ spheres, NPT MD & 3 & 0.79 & - & 2.45E-2 & 4.71E-2 \\
      3DSqW \cite{ParmarEtal2017AG}& 50:50 mixture, square well interaction & 3 & 0.89 &	1.2:1 & 9.44E-3 & 1.43E-2\\
      3DHS \cite{2007AngelaniFoffi} & 50:50 mixture of hard spheres & 3 & 0.91 & 1.2:1 & 5.75E-2 & 7.21E-2\\
      3DBKS \cite{carre2008new}& BKS model of silica & 3 & 0.94 & 1.08:1 & 1.35E-4 & 1.68E-4\\
      3DCHIK \cite{carre2008new} & CHIK model of silica & 3 & 0.94 & 1.08:1 & 1.06E-4 & 1.35E-4\\
      3D-Vitreloy4 \cite{fielitz1999} & Bulk metallic glass (Fe:Be) & 3 & 0.52& 1.56:1.12 & 1.70E-18 & 3.44E-18 \\
      3D-Vitreloy4  & Bulk metallic glass (Fe:B)  & 3 & 0.82& 1.56:0.87 & 5.88E-18 & 1.20E-17 \\
      3D-Vitreloy4  & Bulk metallic glass (Fe:Co) & 3 & 0.94& 1.56:1.52 & 5.88E-18 & 1.07E-17   \\
      3D-Vitreloy1 \cite{fielitz1999}  & Bulk metallic glass (Fe:Be)& 3 & 0.59 & 1.56:1.12 & 1.85E-20 & 1.46E-19  \\
      3D-Vitreloy1  & Bulk metallic glass (Fe:B) & 3 & 0.88 & 1.56:0.87 & 6.29E-19 & 2.34E-18  \\
      3D-D1 \cite{macht2001} & Bulk metallic glass (Fe:B)& 3 & 0.88  & 1.56:0.87 & 3.23E-19 & 7.05E-19 \\
      3D-D2 \cite{macht2001} & Bulk metallic glass (Fe:B)& 3 & 0.93  & 1.56:0.87 & 2.08E-19 & 1.65E-18  \\
      3DLJsoft \cite{SenguptaEtal2011} & 80:20 mixture of modified LJ (12,11) potential   & 3 & 0.87 &	1:0.88 & 3.24E-2 & 5.44E-2 \\
      3DKA \cite{SenguptaEtal2011,SenguptaEtal2013}& 80:20 mixture of Kob-Andersen interaction & 3 & 0.91 &	1:0.88 & 2.03E-1 & 2.90E-1\\
      3DLJsoft  & 80:20 mixture of modified LJ (8,5) potential   & 3 & 0.96 &	1:0.88 & 5.73E-2 & 7.76E-2 \\
      2DKA \cite{SenguptaEtal2013SEB} & 80:20 mixture, Kob-Andersen interaction & 2 & 0.91 &	1:0.88 & 9.01E-3 & 1.18E-2 \\
      2DR12 \cite{perera1999relaxation} & 50:50 mixture, repulsive $r^{-12}$ power law interaction & 2 & 0.96 & 1.4:1 &7.35E-2 &7.89E-2\\ 
      2DR10 \cite{SenguptaEtal2013SEB}& 50:50 mixture, repulsive, $r^{-10}$ power law interaction & 2 & 0.96 & 1.4:1 & 6.54E-2 & 7.64E-2\\
      2DMKA \cite{SenguptaEtal2013SEB}& 65:35 mixture, Kob-Andersen interaction & 2 & 0.99 &	1:0.88 & 2.26E-2 & 2.31E-2             \\
      4DKA \cite{SenguptaEtal2013SEB}& 80:20 mixture of Kob-Andersen interaction & 4 & 0.80 &	1:0.88 & 3.65E-2 & 7.96E-2             \\
      \hline
    \end{tabular}
  \end{center}
\end{table*}

\begin{figure*}[t]
  \includegraphics[scale=0.18]{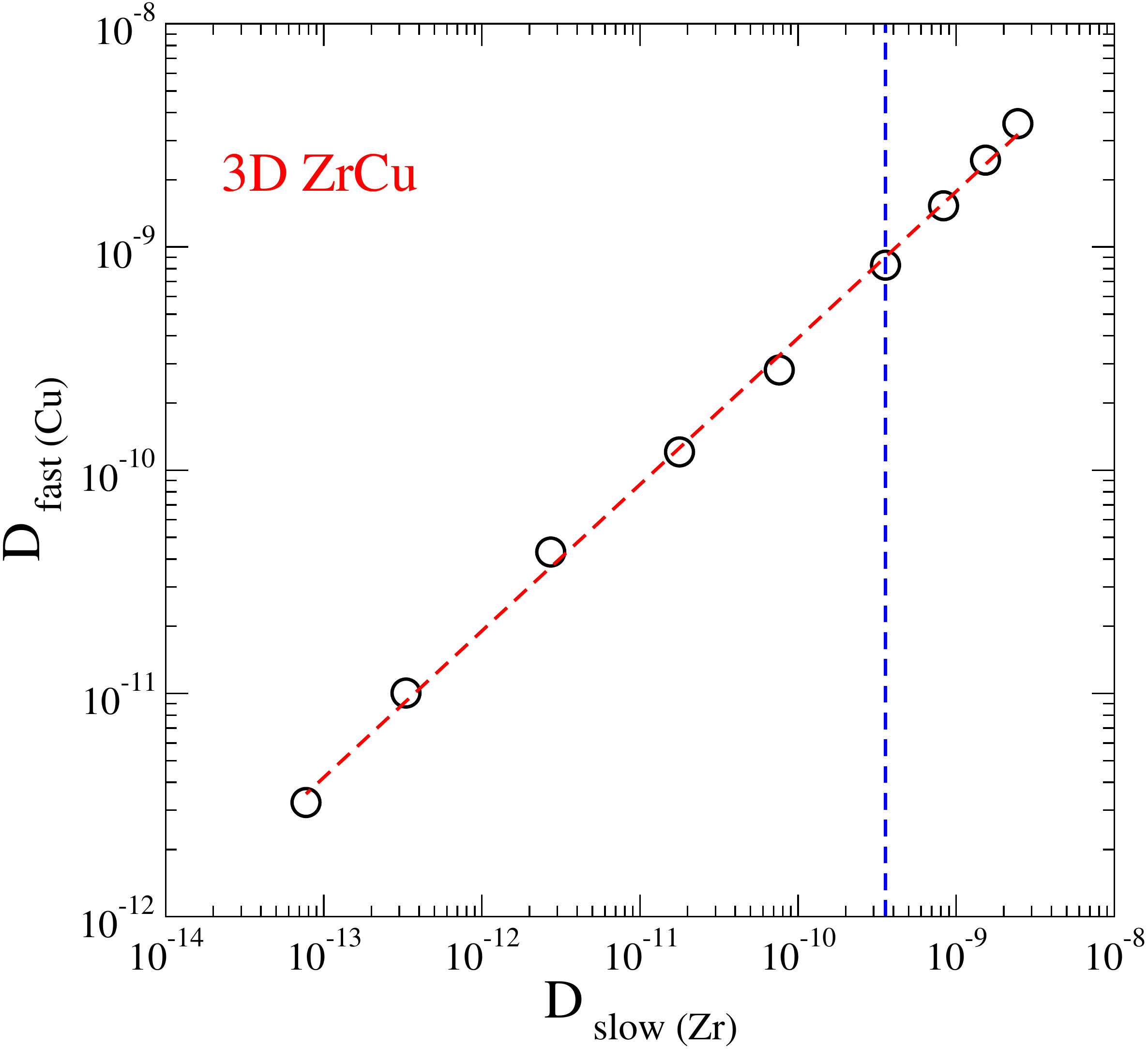}
  \includegraphics[scale=0.18]{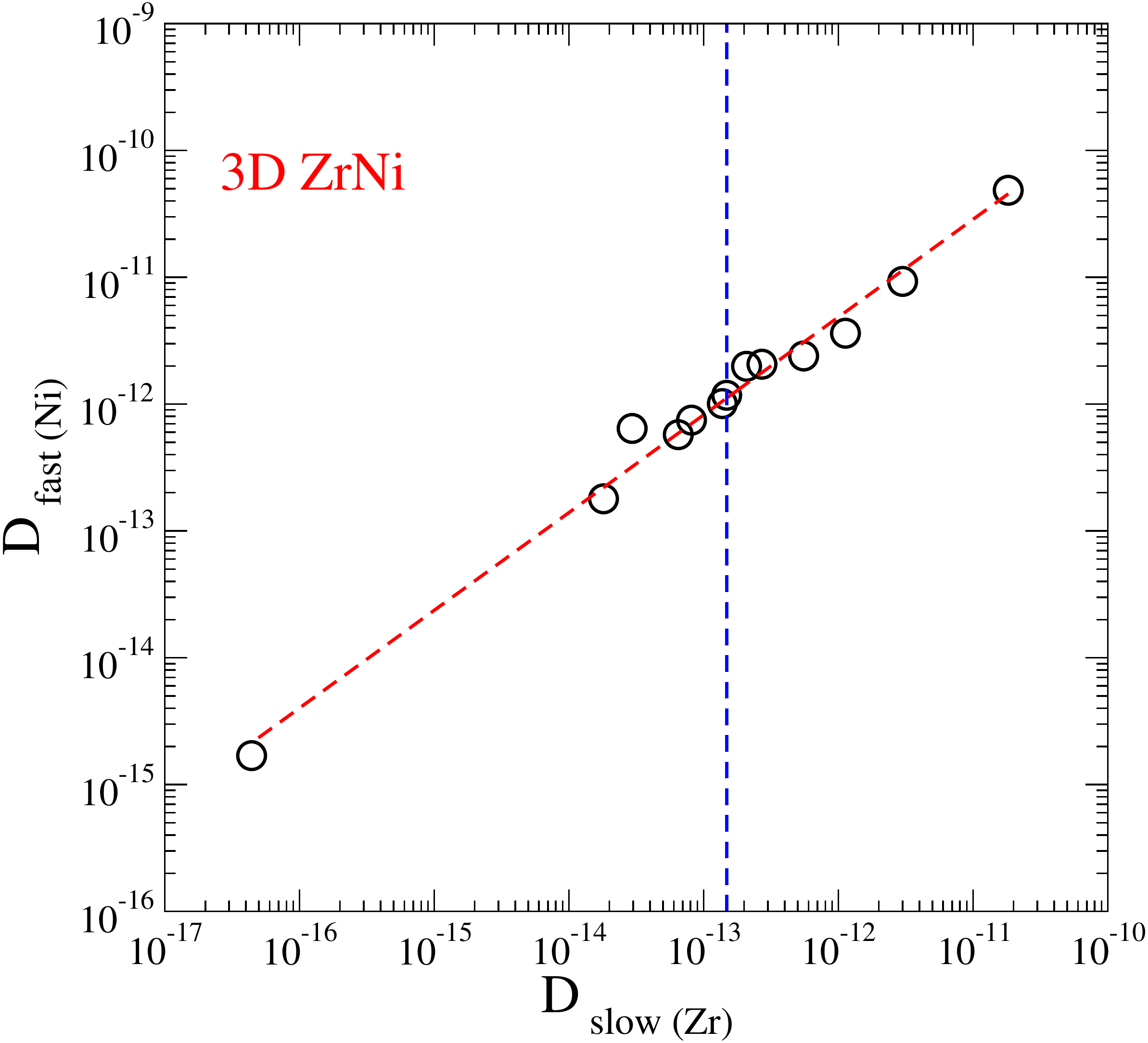}
  \includegraphics[scale=0.18]{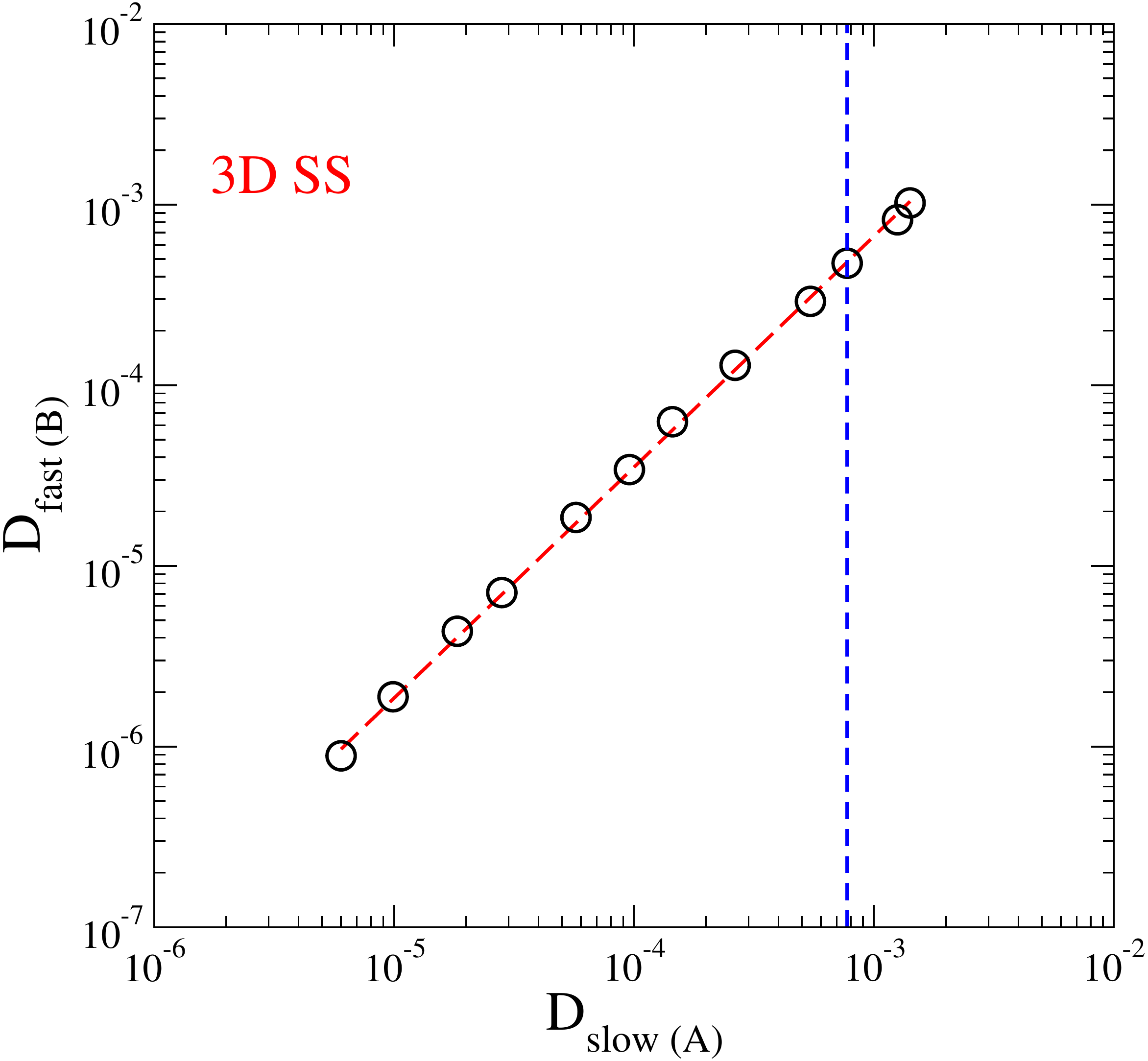}
  \includegraphics[scale=0.18]{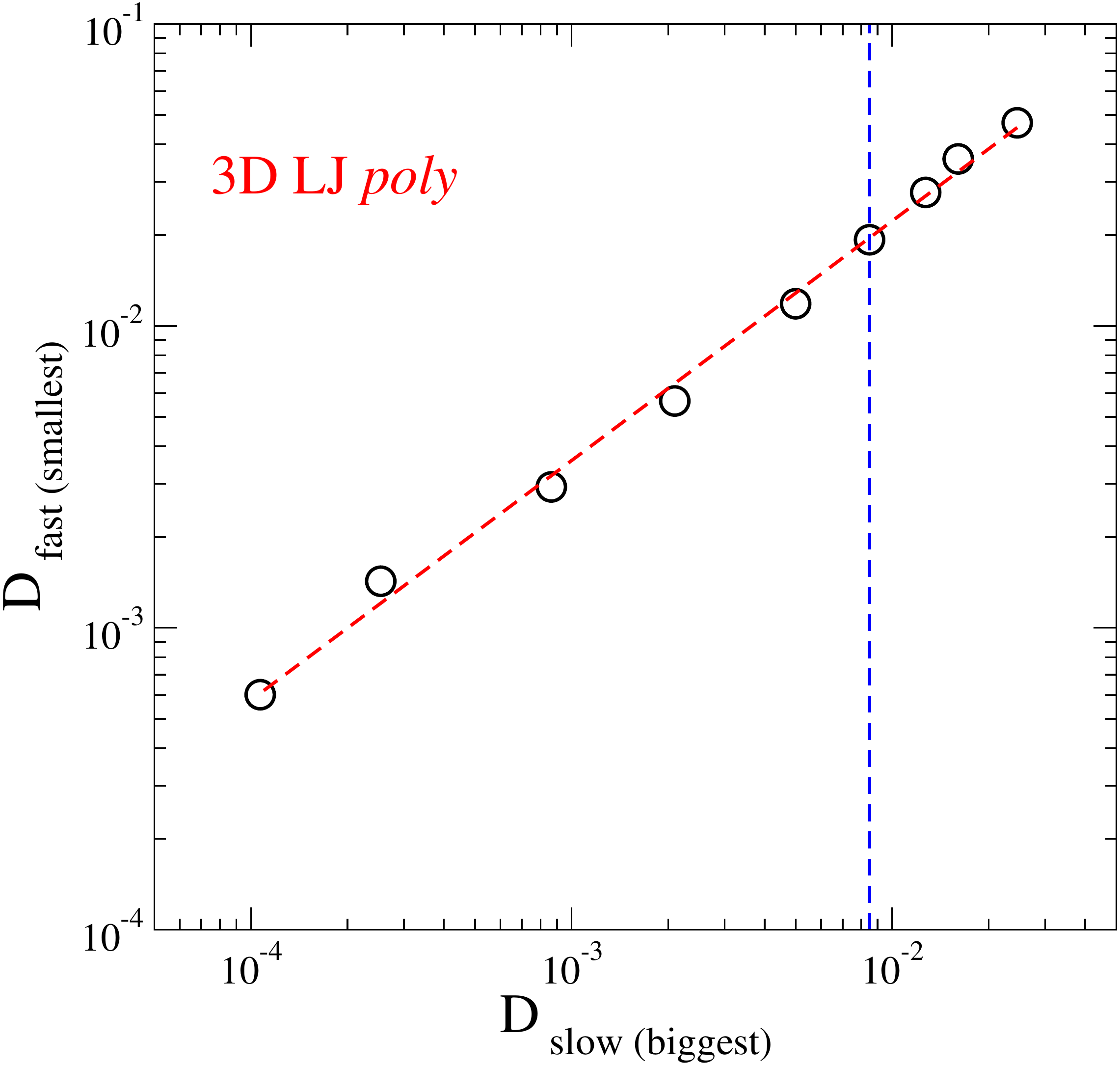}\\
  \vspace{5mm}
  \includegraphics[scale=0.18]{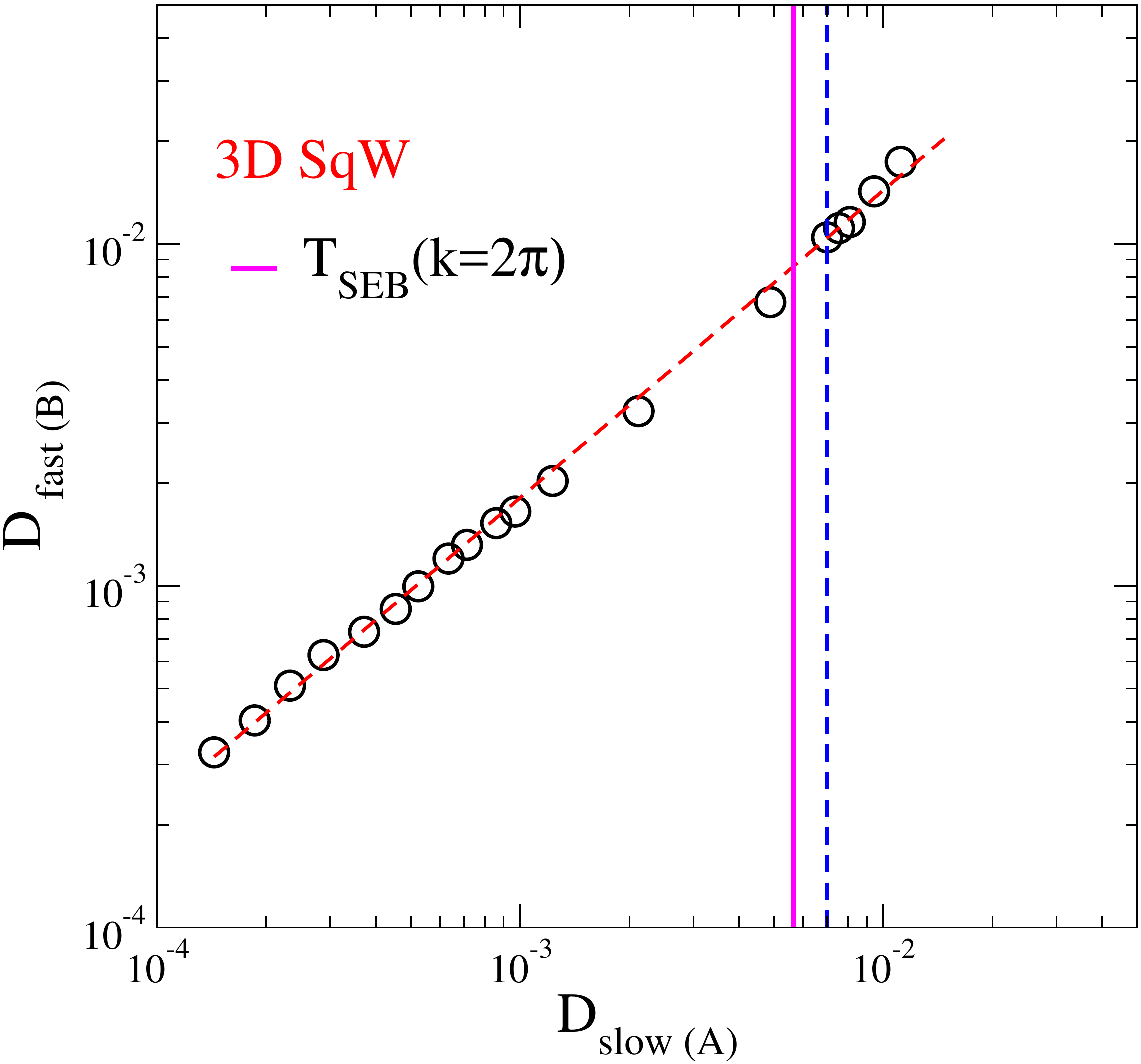}
  \includegraphics[scale=0.18]{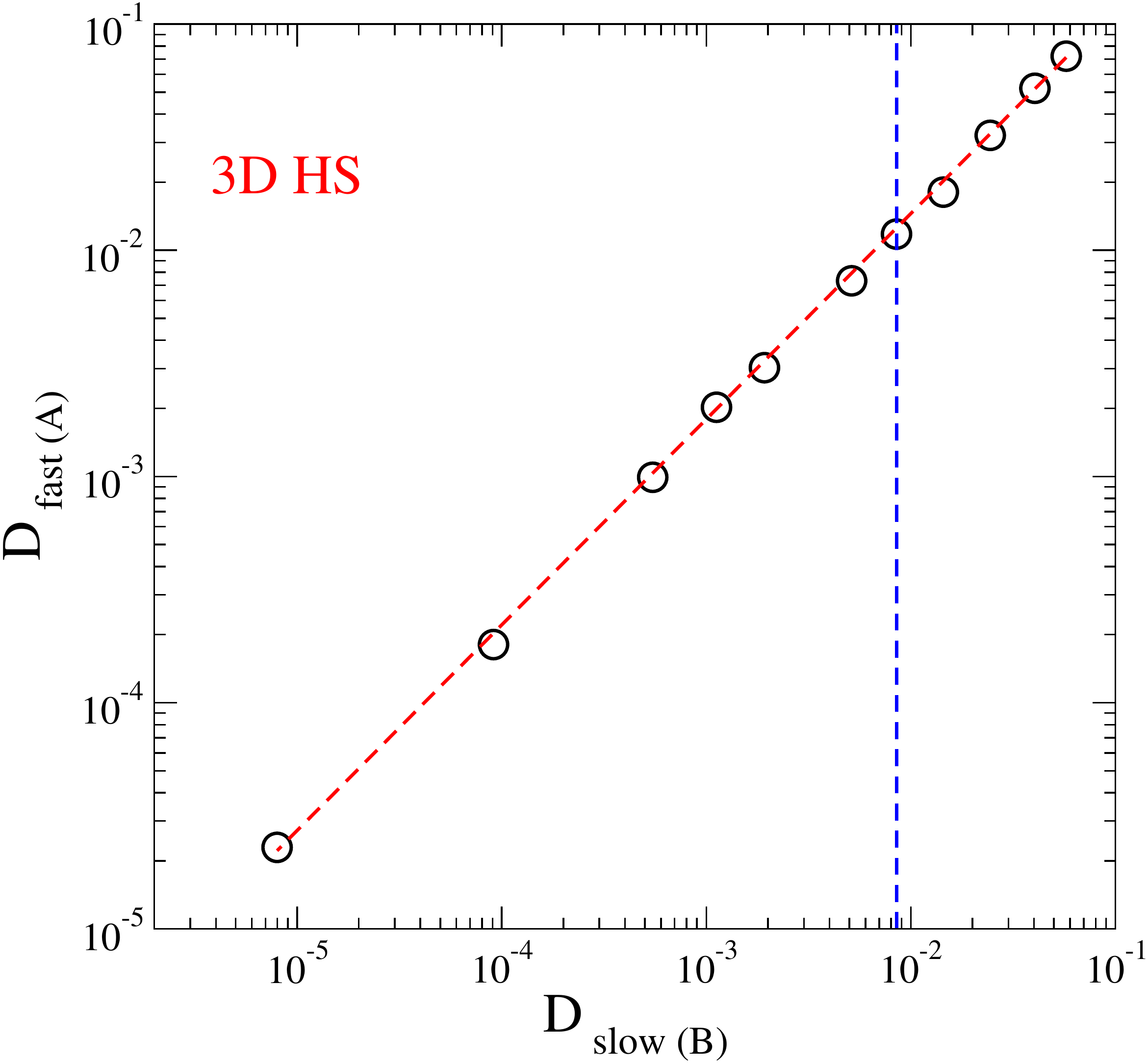}
  \includegraphics[scale=0.18]{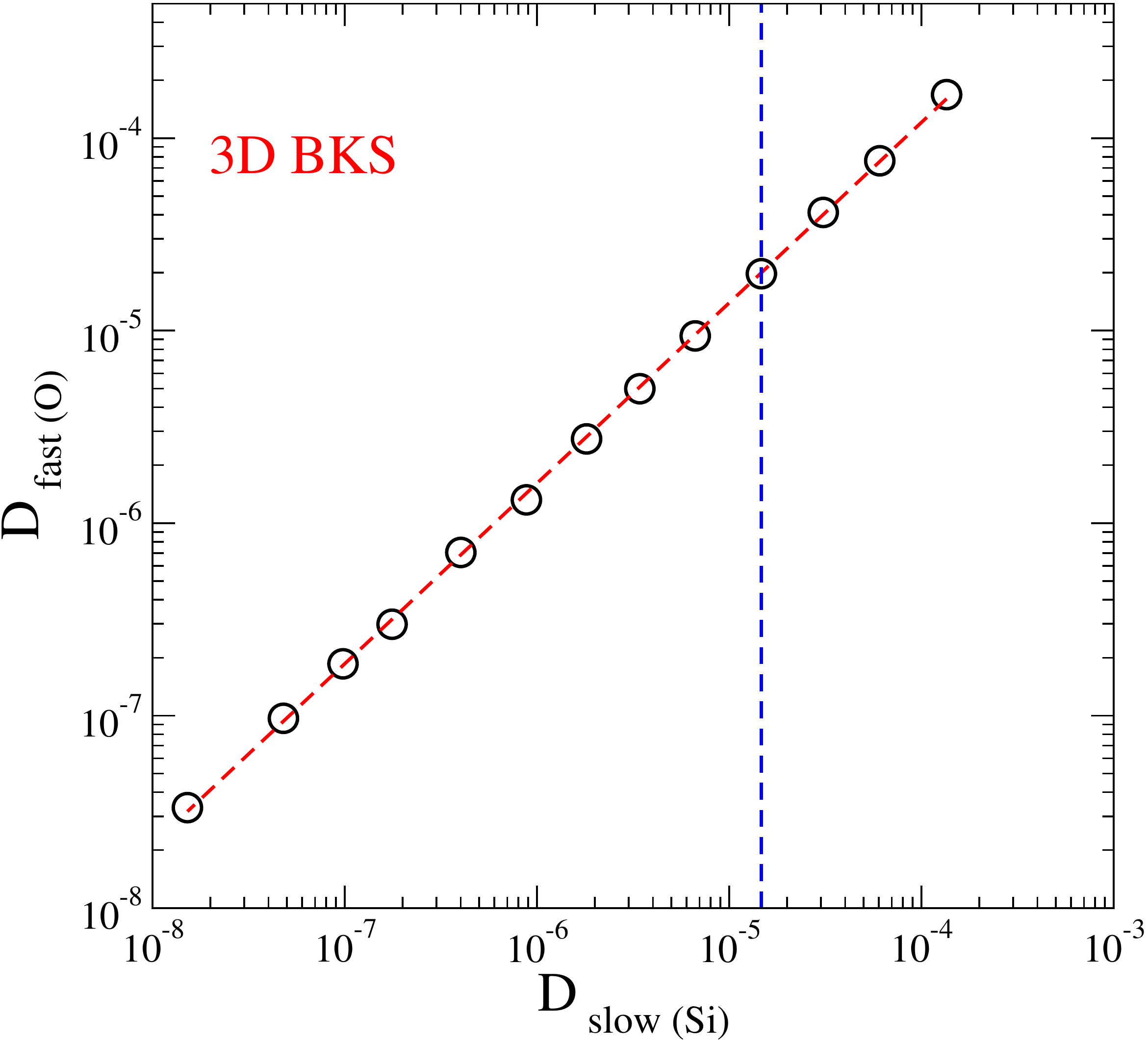}
  \includegraphics[scale=0.18]{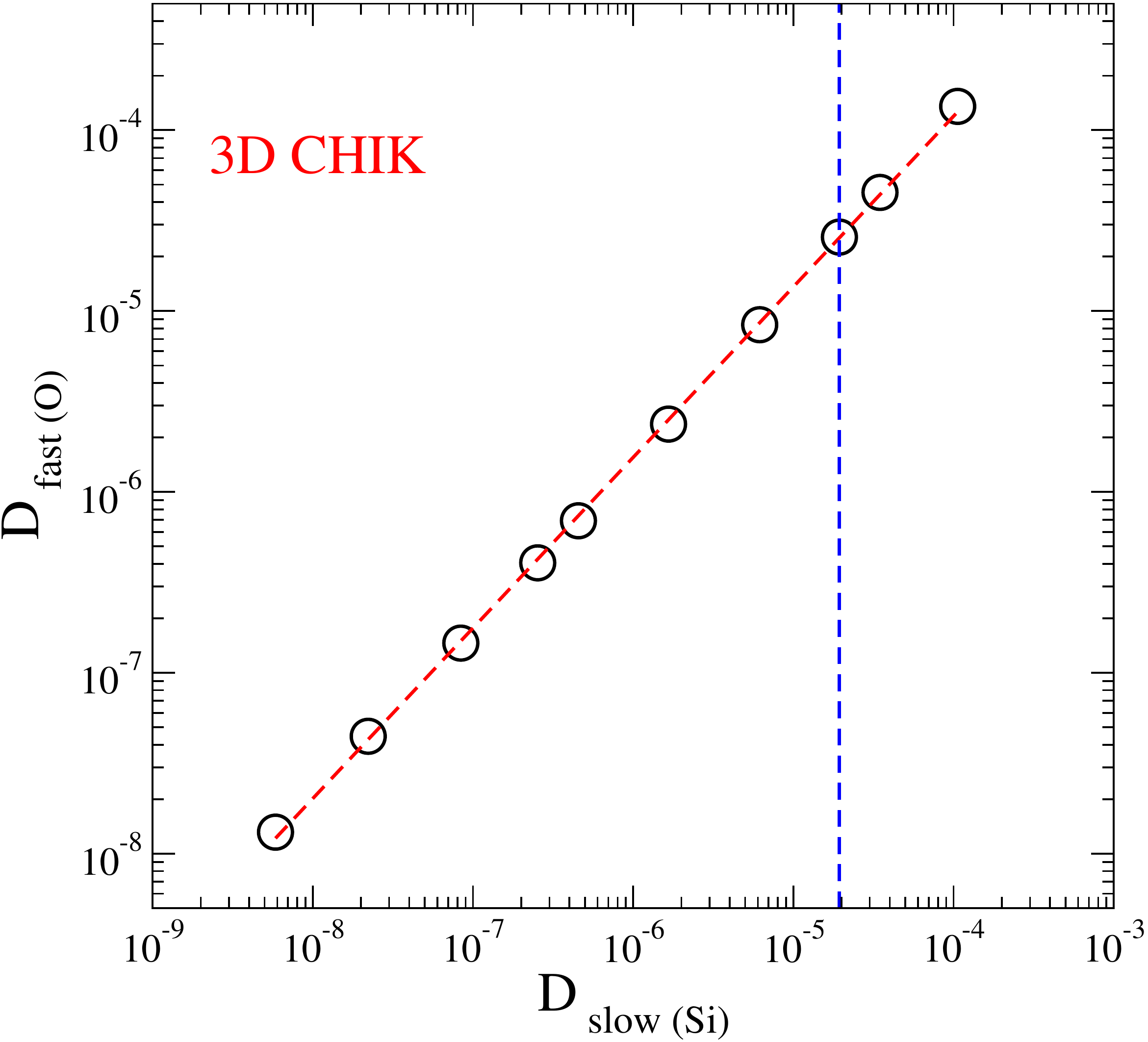}\\
  \caption{Power law dependence of diffusion coefficient in several
   \textcolor{black}{simulation} model glass-formers in 3 dimensions. By
   convention, the slow component is plotted on X axis and the fast
   component on Y-axis. Vertical dashed lines indicate the onset
   temperature/density and continuous lines (for 3DSqW \cite{ParmarEtal2017AG}) indicate the Stokes-Einstein breakdown temperature estimated from the structural relaxation times.} 
 \label{3Dsim}
\end{figure*}

\begin{figure*}[htp]
\centering
\includegraphics[scale=0.18]{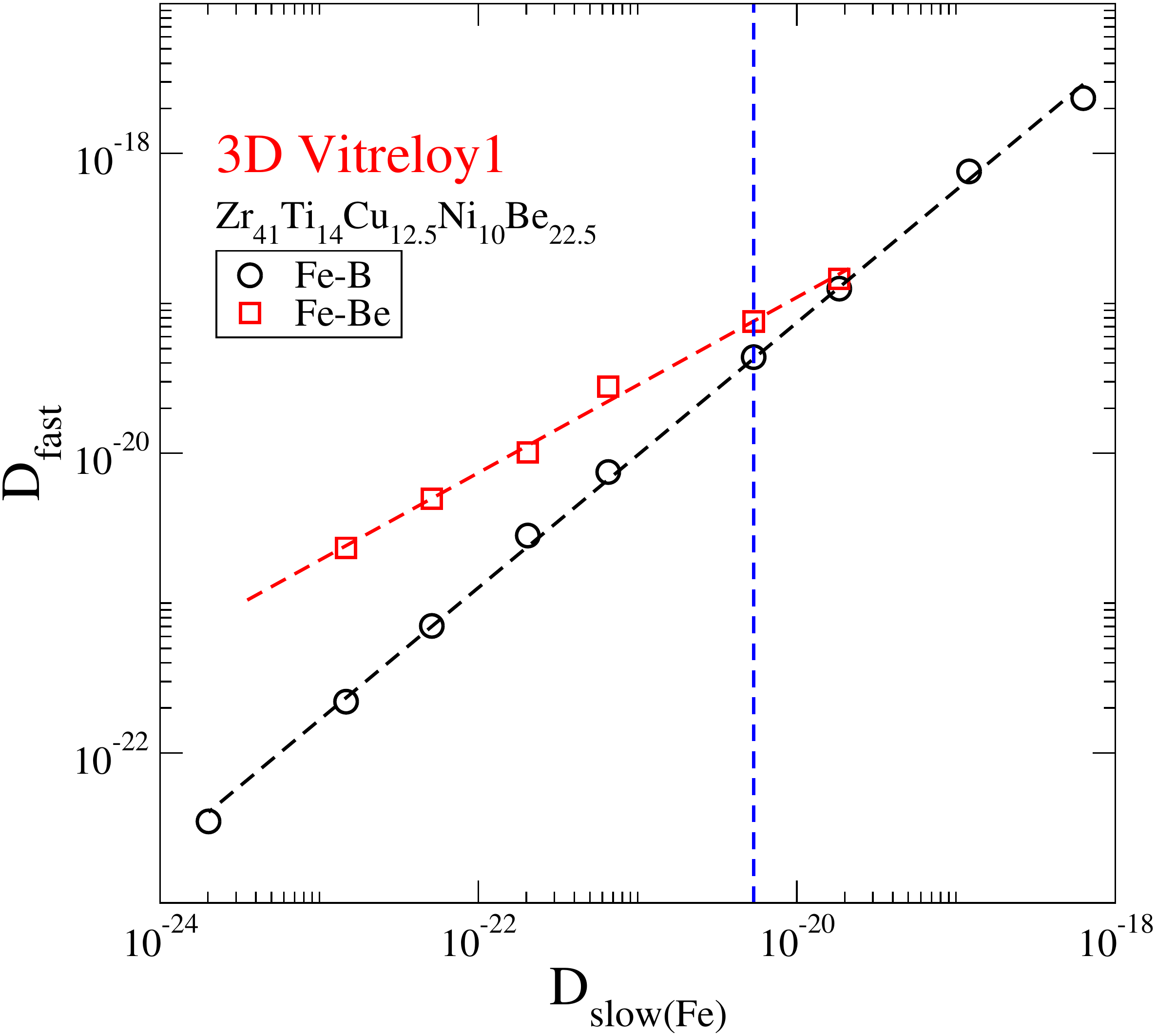}
\includegraphics[scale=0.18]{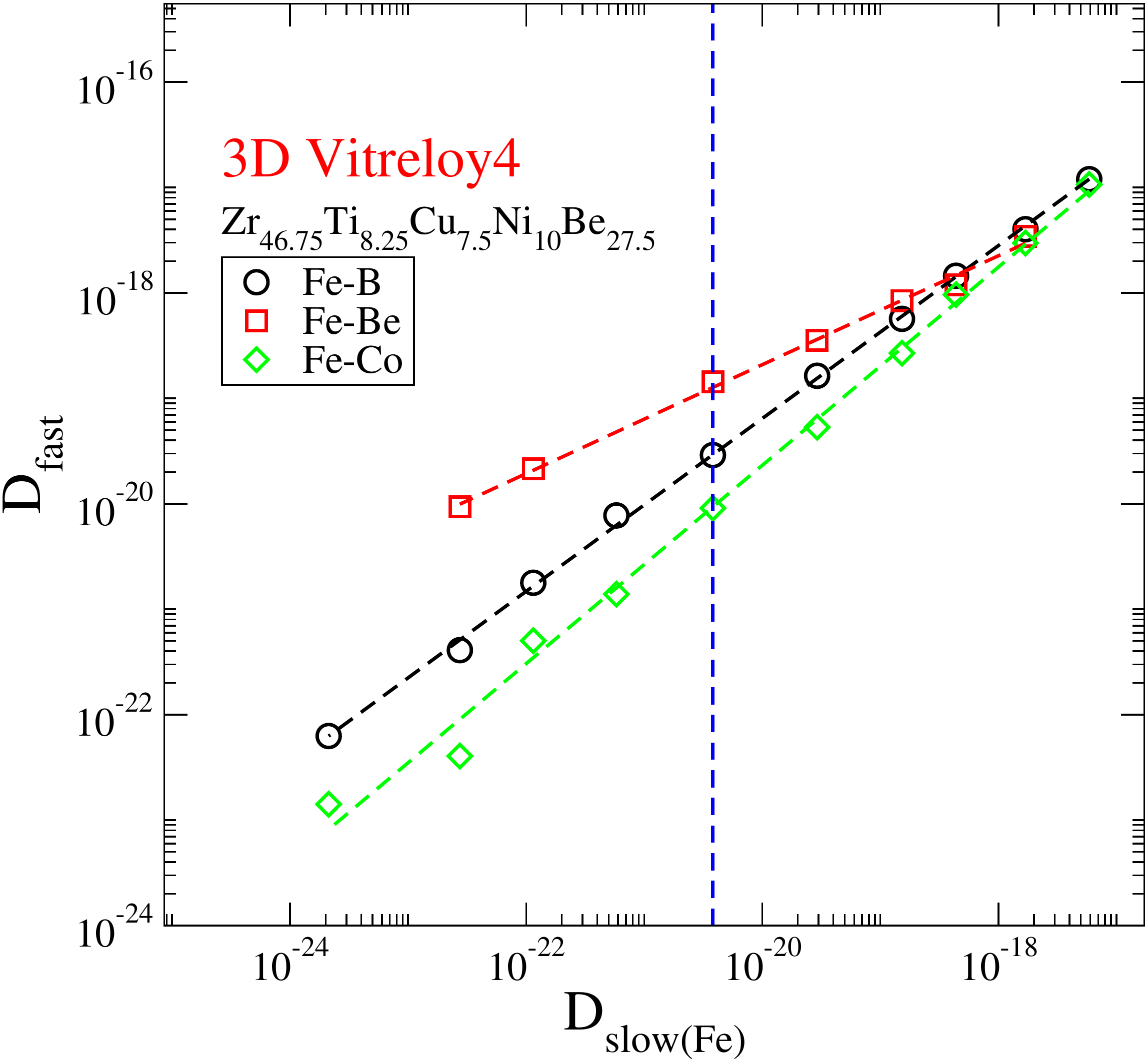}
\includegraphics[scale=0.18]{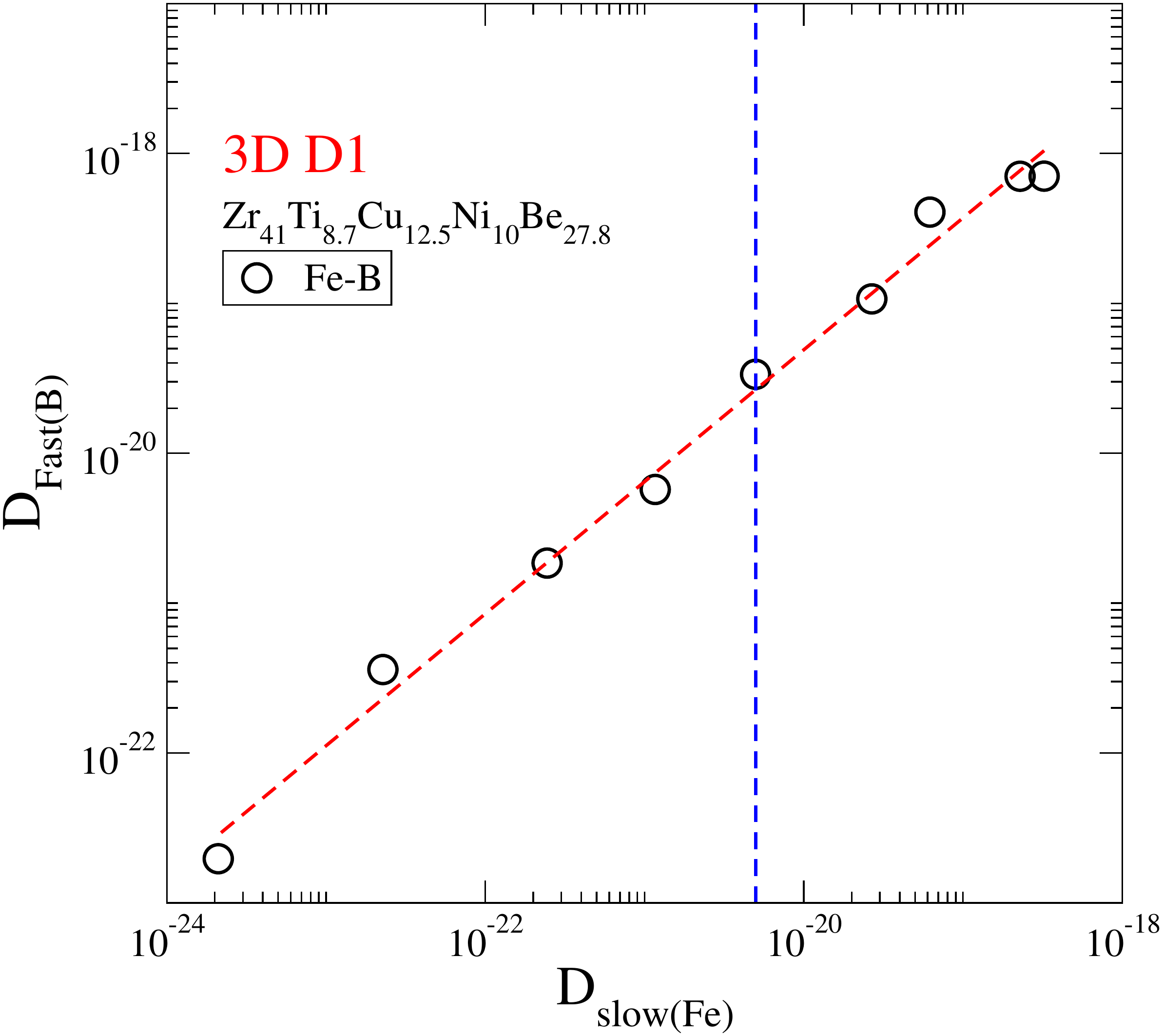}
\includegraphics[scale=0.18]{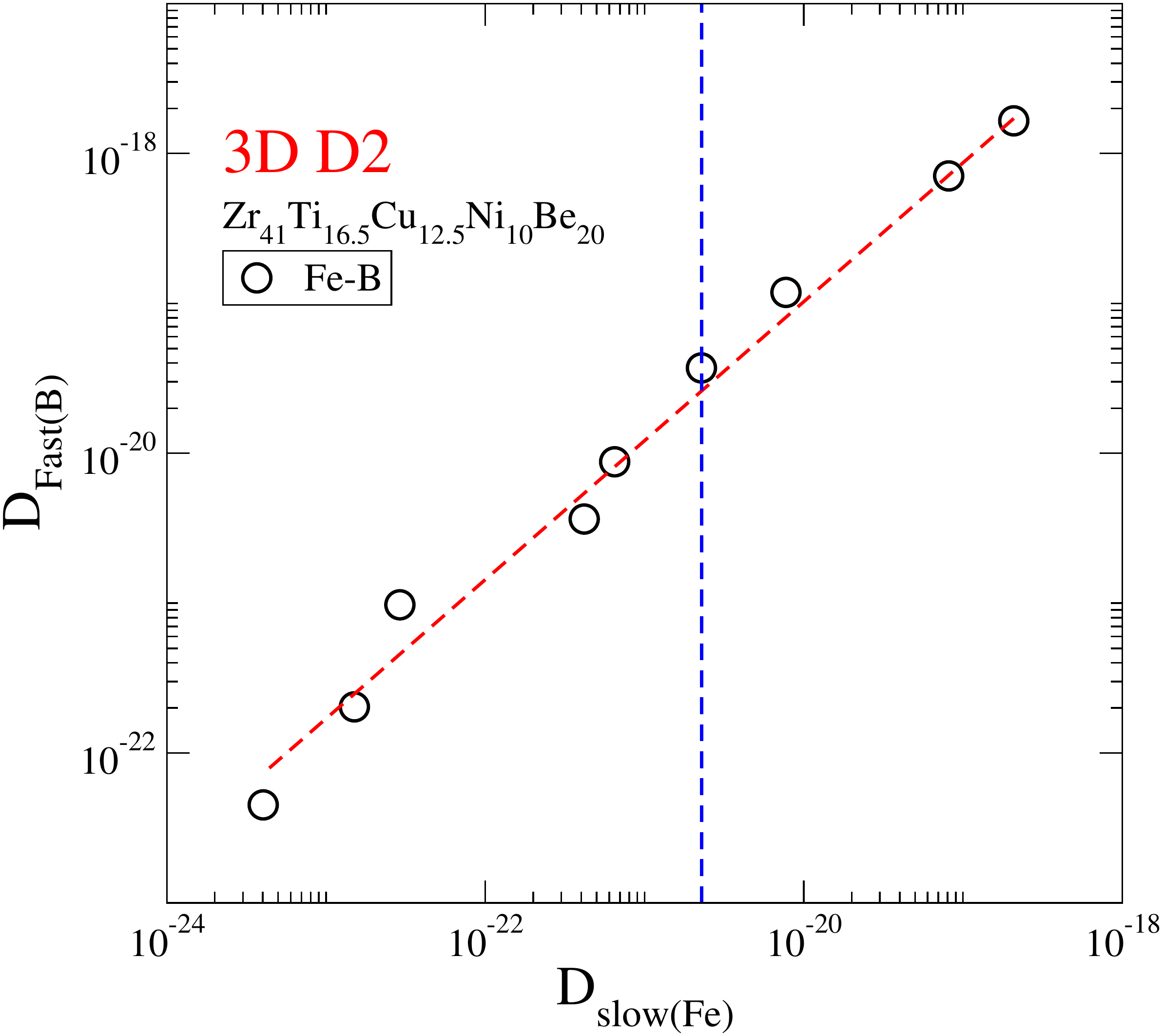}
\caption{Power law dependence of diffusion coefficient in several
  \textcolor{black}{experimental model} glass forming
  liquids. Vertical lines indicate the onset temperature.} 
\label{3Dexpt}
\end{figure*}

\section{Diffusion Cofficients for Multi-Component Glass Forming Liquids}
\label{sec:models}
In this section, we describe the details of the glass-formers
surveyed. The data are shown in Figs. \ref{3Dsim}-\ref{2D} and the
relevant parameters are tabulated in Table \ref{tab:table1}. 
We report the data for diffusion coefficients over a wide range of the temperatures or densities across the onset temperature. The onset temperature 
has been estimated by considering the deviation from the high
temperature Arrhenius behaviour. Vertical lines in the figures mark
the diffusion coefficient value  of the bigger particles at the onset
temperature or the breakdown of the Stokes-Einstein relation. \textcolor{black}{When taken from literature,} the source
of the data in each case is indicated by appropriate references
\textcolor{black}{after the acronym}.

\subsection{Simulation}

\textcolor{black}{
\begin{enumerate}
\item 3DZrCu \cite{kluge2004diffusion}- Binary mixture of the metallic
  glass $Cu_{33}Zr_{67}$ modelled by embedded atom potential and simulated by $NPT$ molecular dynamics ($MD$) simulations. 
\item 3DZrNi \cite{teichler2001structural}- 50:50 binary mixture of
  $Zr_{0.5}Ni_{0.5}$ metallic glass simulated by $NPT$ $MD$. 
\item 3DSS- 50:50 binary mixture of soft
  spheres in $3$ dimensions (model details as in \cite{berthier2009glass}). We have performed $MD$ simulations in the $NVT$ ensemble have been performed for a temperature range \textcolor{black}{in units of
    ${\epsilon \over k_B}$, where $\epsilon$ is the strength of the
    potential} is [0.00165, 0.00390], and the packing fraction is $\phi=0.75$. The packing fraction is defined as $\phi=(\sum_{i}\rho_{i}\sigma_{ii}^{3})\pi/6$, where $i$ represents particle type, $\rho_{i}$ is the number density and $\sigma_{ii}$ the diameter of particle type $i$.
\item 3DLJpoly \cite{murarka2003diffusion}: Polydispersed $LJ$ system studied in the $NPT$ ensemble  at
  pressure $P=10 {\epsilon \over \langle \sigma \rangle^3}$.
\item 3DSqW- 50:50 binary mixture of particles
  interacting {\it via} an attractive, square well potential\cite{saika2004effect}. The
  temperature range is \textcolor{black}{$[0.31,10] {\epsilon \over
      k_B}$ at constant density $\rho=0.77$, where $\epsilon$ is the strength of the potential}
  \cite{ParmarEtal2017AG}.
\item 3DHS \cite{2007AngelaniFoffi}- 50:50 binary mixture of hard
  spheres. The range of packing fractions [0.425, 0.58] studied at constant temperature $T=1$.
\item 3DBKS- BKS model for silica in $3$ dimensions
  \cite{carre2008new}. The temperature range $[2750, 6100] K$ studied at the density $2.37g/cm^{3}$.
\item 3DCHIK- CHIK model for silica in $3$ dimensions
  \cite{carre2008new}. The temperature range $[2440, 5200] K$ studied at the density $2.20g/cm^{3}$.
\item \textcolor{black}{3DLJsoft- 80:20 binary mixture of
    particles interacting {\it via} modified Lennard-Jones (LJ)
    models in 3 dimensions. The softness is varied by tuning the
    exponents of the repulsive ($q$) and attractive ($p$) part of the
    LJ potential \cite{SenguptaEtal2011}. The three models we consider have exponents
    ($q=12,p=11$), ($q=12,p=6$), and ($q=8,p=5$) respectively. The
    number density for all models is $1.20 \, \sigma_{AA}^{-3}$.}
\item 3DKA- 80:20 binary mixture of particles interacting {\it via}
  the Kob-Andersen potential \cite{kob1995testing} in 3
  dimensions. The number density is $1.20 \,\sigma_{AA}^{-3}$, and the 
  temperature range is $[0.46,6.0]\, {\epsilon_{AA} \over k_B}$.
\item 2DKA- 80:20 binary mixture of particles interacting {\it via}
  the standard Kob-Andersen potential in 2 dimensions. The temperature
  range is $[0.93, 2.00]\,{\epsilon_{AA} \over k_B}$ studied at the number density $1.20\, \sigma_{AA}^{-2}$.
\item 2DMKA- 65:35 binary mixture of particles interacting {\it via}
  the Kob-Andersen potential in 2 dimensions. The temperature
  range  is $[0.45, 1.50]\, {\epsilon_{AA} \over k_B}$, studied at the number density $1.20\, \sigma_{AA}^{-3}$.
\item 2DR10- 50:50 binary mixture of
  particles interacting via purely repulsive inverse power law
  $r^{-10}$ potential in $2$ dimensions
  \cite{karmakar2010predicting}. The temperature range is [0.52, 2.00], studied at the number density $0.85\, \sigma_{AA}^{-2}$.
\item 2DR12 \cite{perera1999relaxation}- 50:50 binary mixture of
  particles interacting via purely repulsive inverse power law
  ($r^{-12}$) potential in $2$ dimensions. The temperature range is $[0.4, 4.0]$, studied {\it via} NPT MD simulations.
 \item 4DKA- 80:20 binary mixture of particles interacting {\it via}
  the standard Kob-Andersen potential in $4$ dimensions at the reduced
  number density $1.6 \,\sigma_{AA}^{-3}$. The temperature
  range is $[0.80, 2.40] {\epsilon_{AA} \over k_B}$
  \cite{SenguptaEtal2013SEB}. 
\end{enumerate}
}

\subsection{Experiment}

\textcolor{black}{
\begin{enumerate}
\item 3D-Vitreloy4 and 3D-Vitreloy1 \cite{fielitz1999}: Five-component
  bulk metallic glass alloy Vitreloy4
  ($Zr_{46.75}Ti_{8.25}Cu_{7.5}Ni_{10}Be_{27.5}$) and Vitreloy1
  ($Zr_{41} Ti_{14} Cu_{12.5} Ni_{10} Be_{22.5}$). The data are
  reported for tracer particles of Boron (B), Iron (Fe), Cobalt (Co)
  and Beryllium (Be).
\item 3D-D1 and 3D-D2 \cite{macht2001}: They are respectively five-component
  bulk metallic glass alloys $Zr_{41} Ti_{8.7} Cu_{12.5} Ni_{10}
  Be_{27.8}$ (D1) and $D2$ $Zr_{41} Ti_{16.5} Cu_{12.5} Ni_{10}
  Be_{20}$ (D2). The data are reported for tracer particles of Boron
  (B) and Iron (Fe).
\end{enumerate}
}

\begin{figure}[htp]
  \centering
  \includegraphics[scale=0.18]{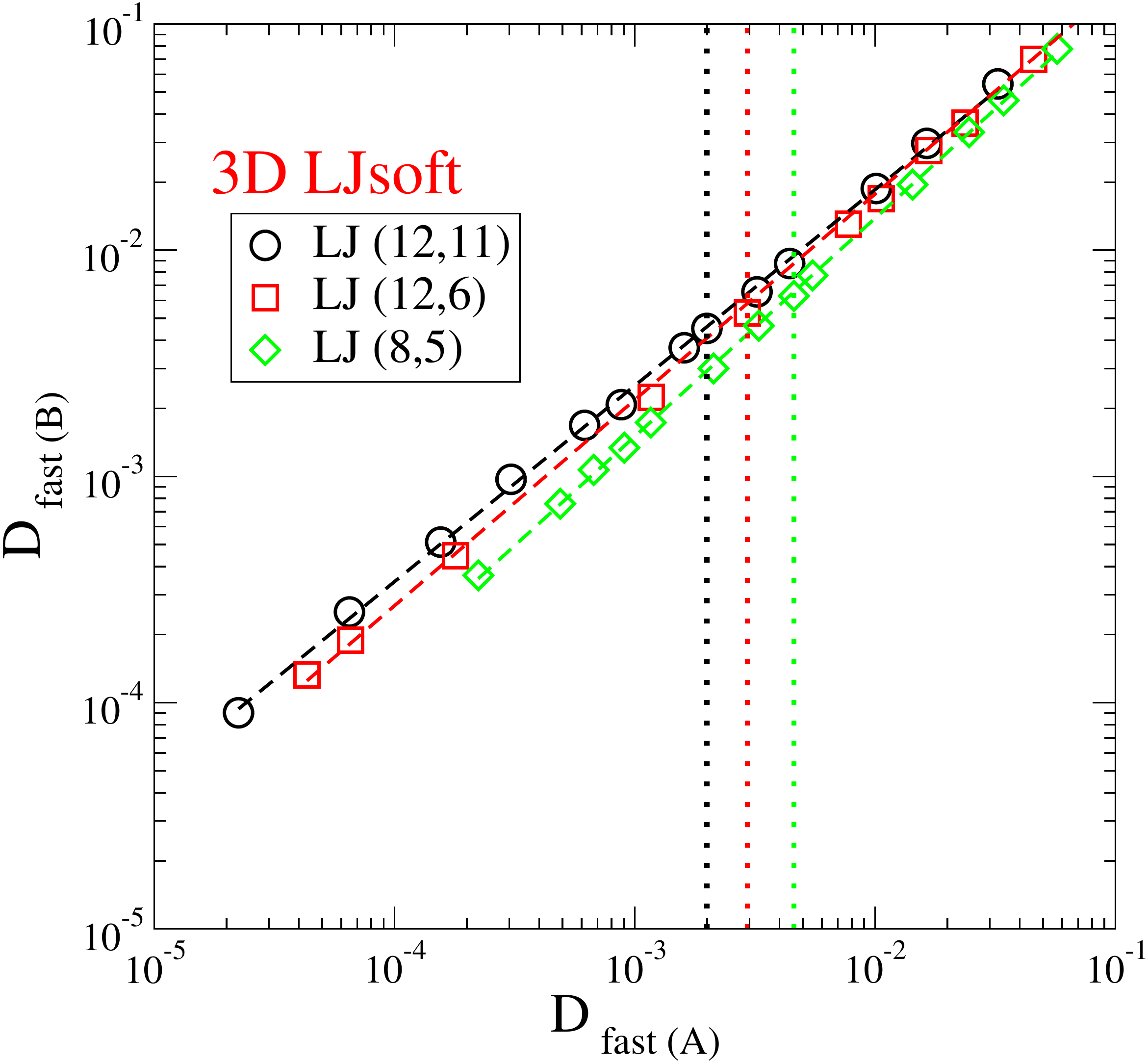}
  \includegraphics[scale=0.18]{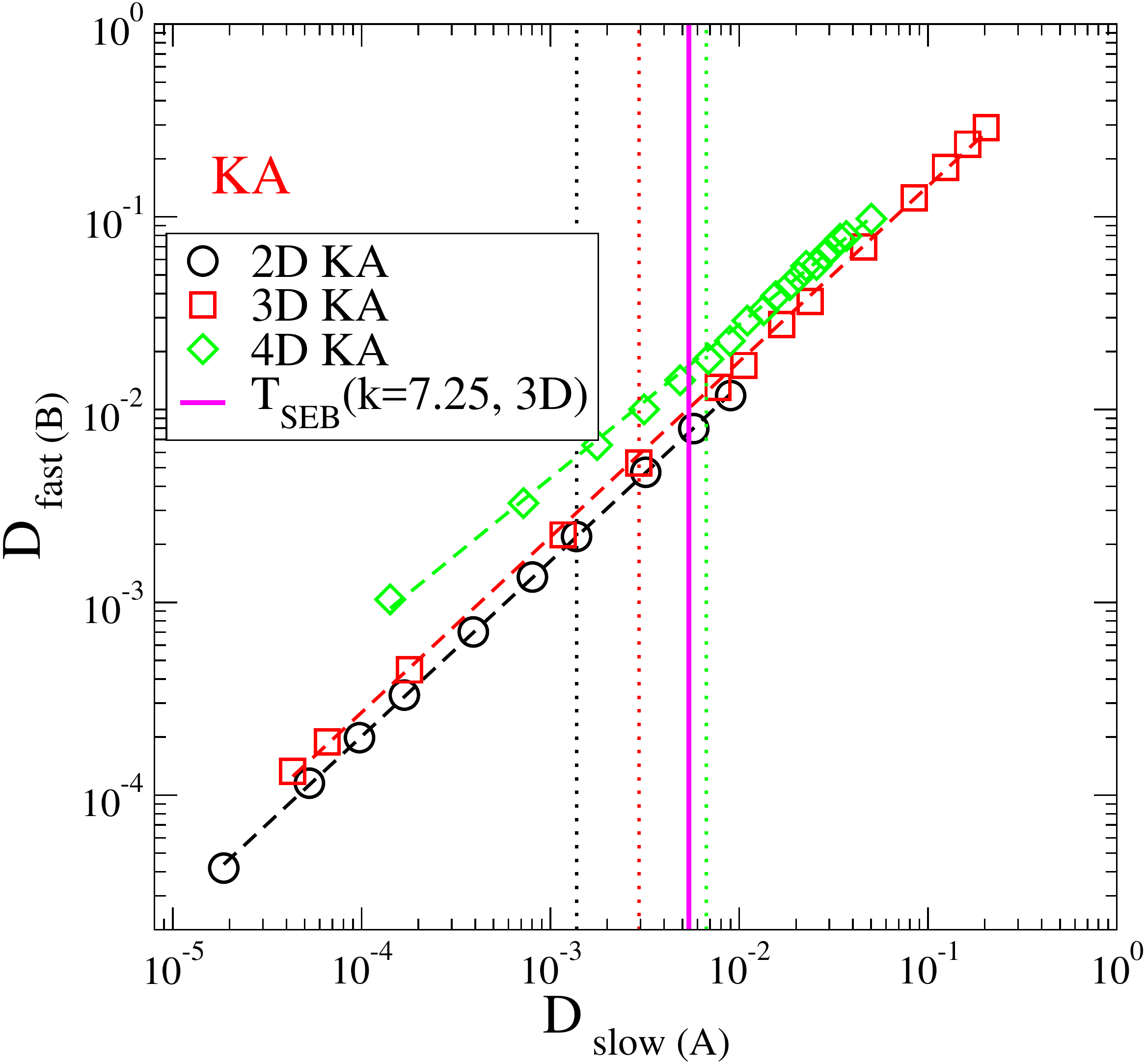}
\caption{\textcolor{black}{Examining the power law relationship among diffusion
  coefficients of different components by systematically
  varying softness (left panel) and the spatial dimension (right
  panel) in Lennard-Jones family of model glass-formers. We see that the power
   law dependence holds in all cases, but the exponent systematically depends
on the model details. Vertical dashed lines indicate the onset
   temperature/density and continuous lines (for 3DKA \cite{ParmarEtal2017AG}) indicate the Stokes-Einstein breakdown temperature estimated from the structural relaxation times.}}
\label{varLJ}
\end{figure}

\begin{figure}[htp]
\centering
\includegraphics[scale=0.18]{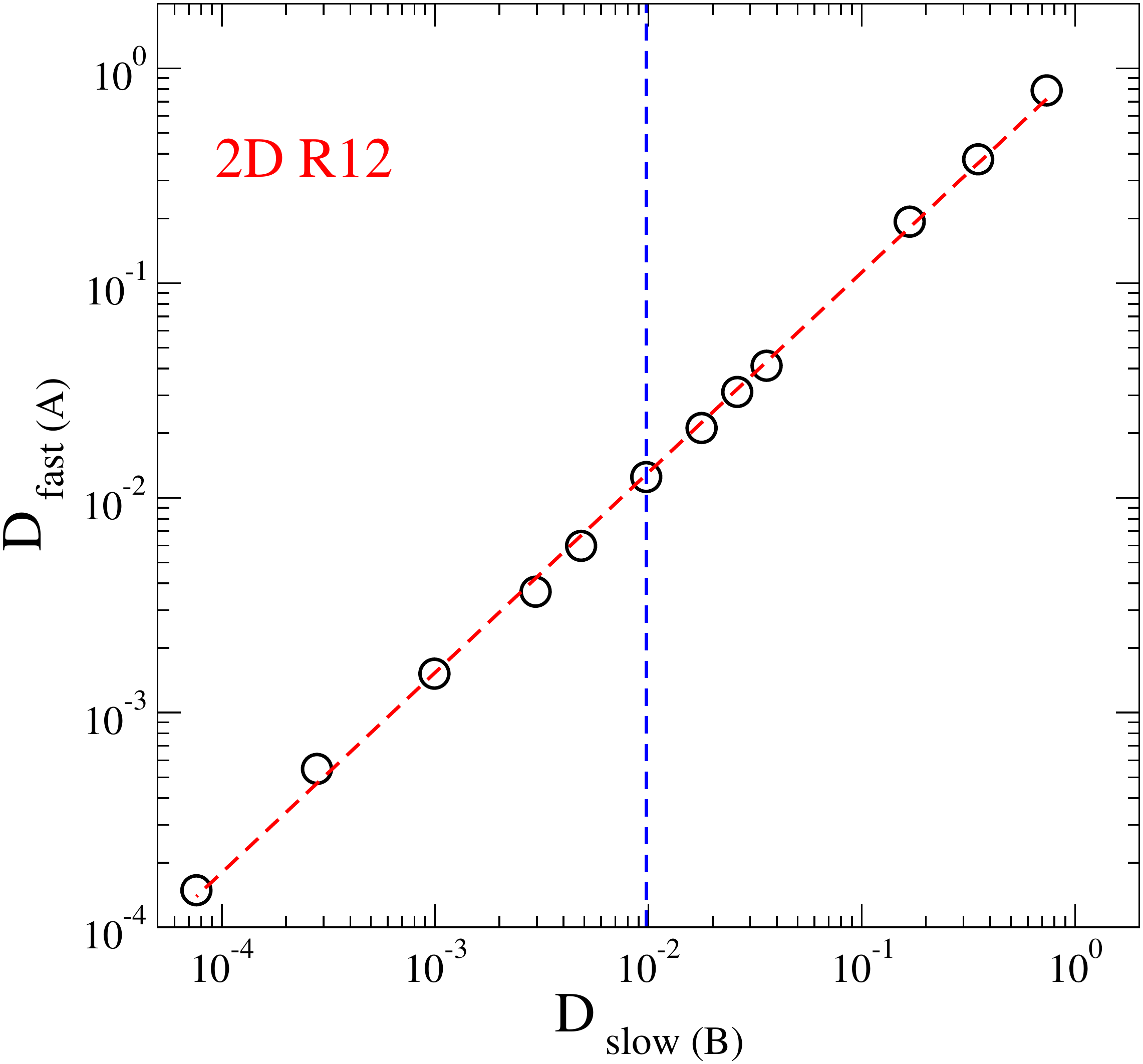}
\includegraphics[scale=0.18]{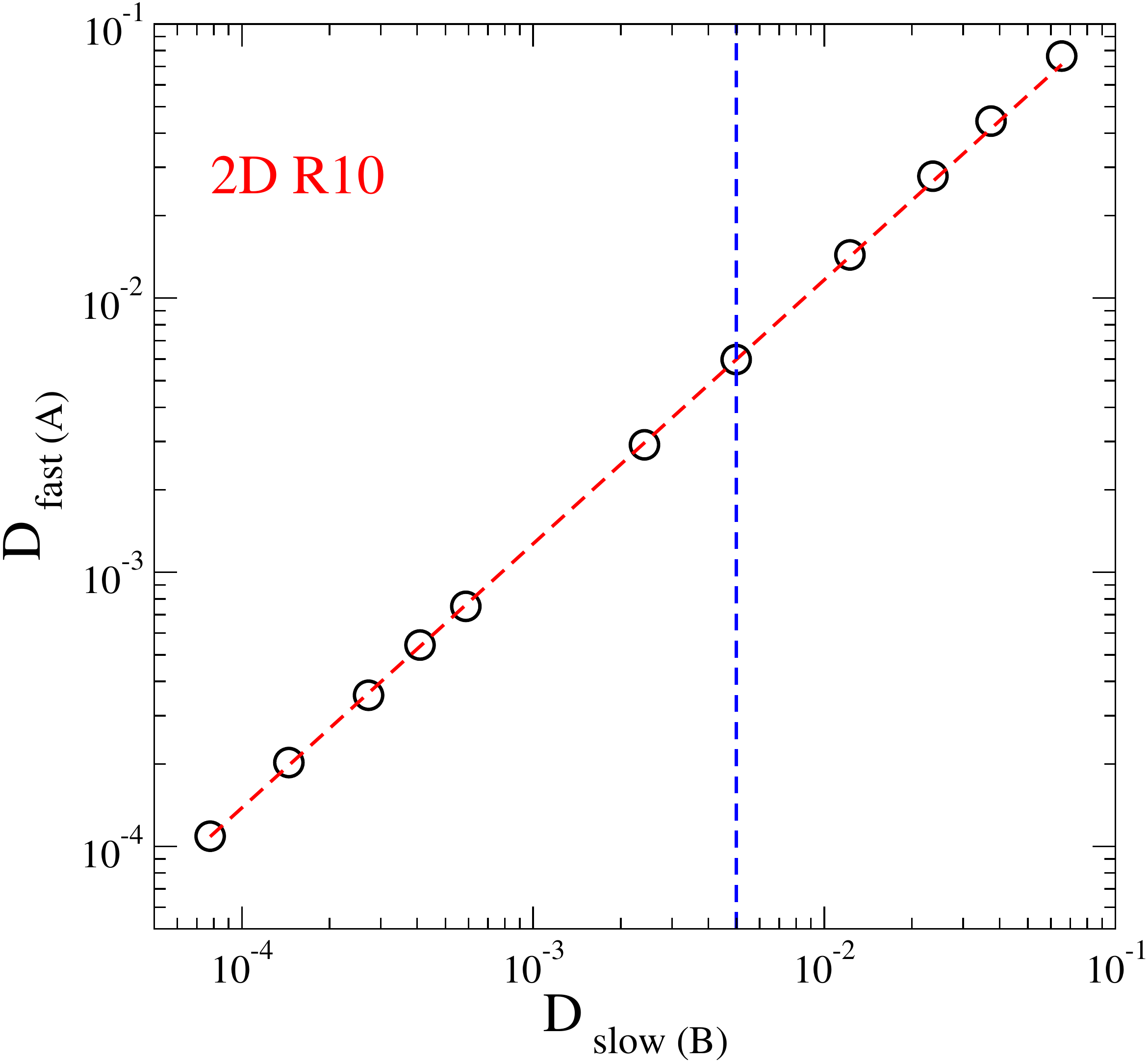}
\includegraphics[scale=0.18]{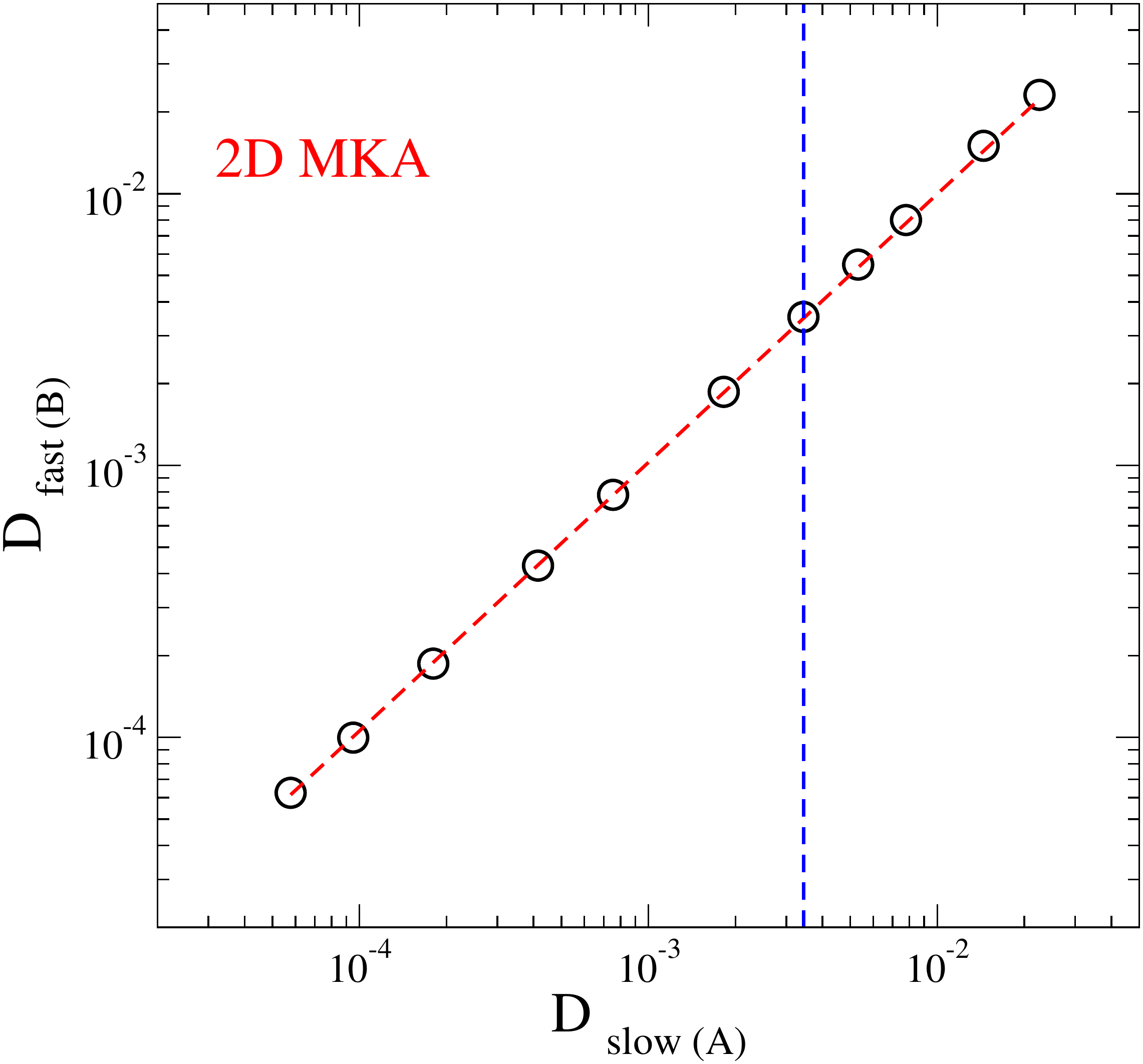}
\caption{Power law dependence of diffusion coefficient in several
  model glass-formers in 2 dimensions. By convention, the slow
  component is plotted on X axis and the fast component on
  Y-axis. Vertical lines indicate the onset temperature.}
\label{2D}
\end{figure}

\begin{figure}[h!]
\centering
\includegraphics[scale=0.28]{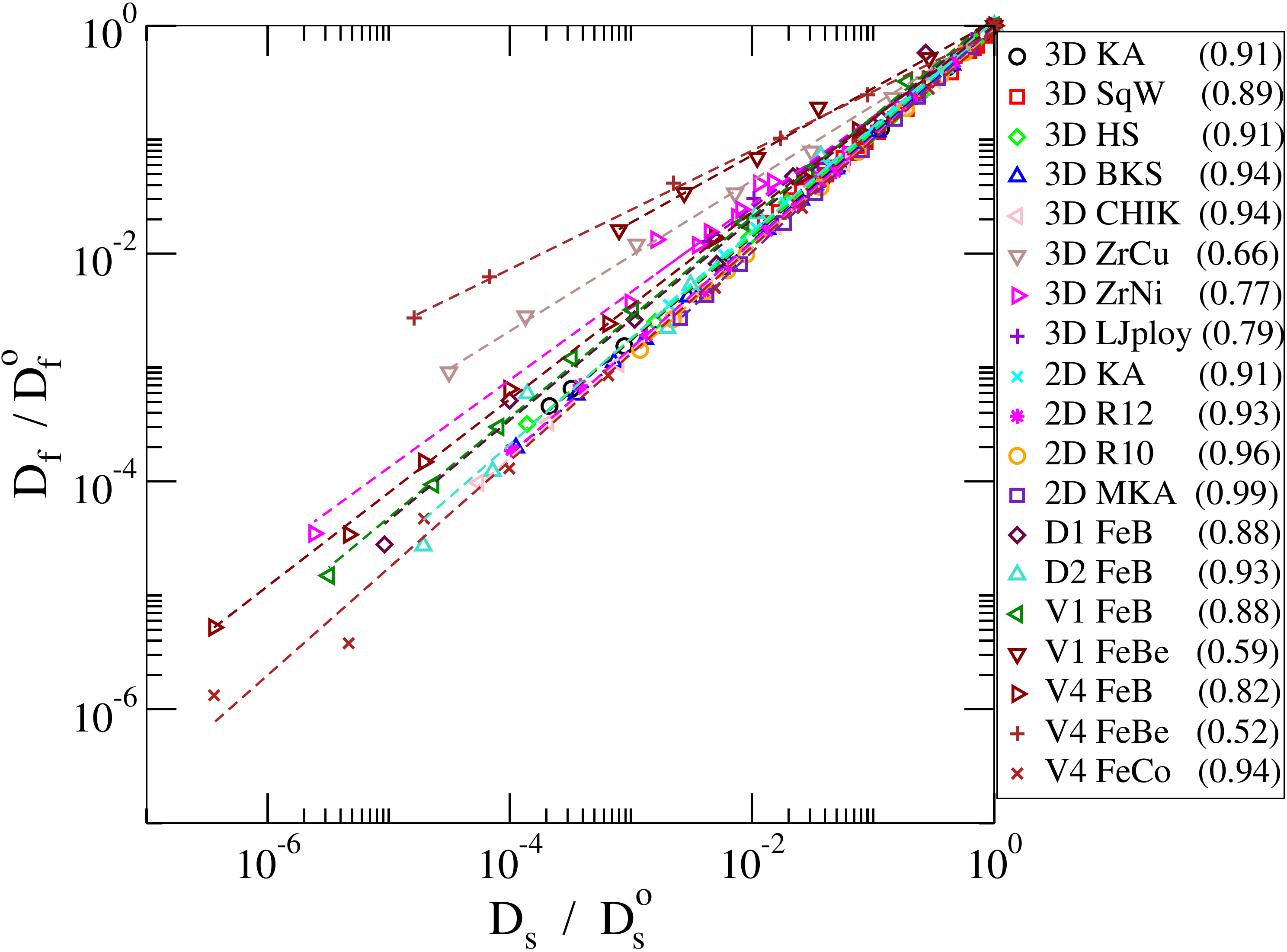}
\caption{Scaled plot of diffusion coefficients for all models across
  dimensions, highlighting the wide range of validity of the power law
  dependence. Indices $s$ and $f$ denote slow and fast components, respectively. Diffusion coefficients are scaled with the corresponding high-temperature diffusivity ($D^{0}$).  
   }
\label{all}
\end{figure}

\section{Discussion}
\label{sec:disc}

Figs. \ref{3Dsim} - \ref{2D} show the data of diffusion coefficients
where the slow component is plotted along the X-axis, and the fast
component is plotted along the Y-axis, by convention. It is clear that in all cases, a fractional power law describes the data well, and very convincingly so in most cases.  We also note that the exponent is generally different
from $1$, with the exponent values being sensitive to the composition and the
interaction potential. \textcolor{black}{This is systematically shown
  in Fig. \ref{varLJ}, where softness of the potential and the spatial
  dimension are varied for a family of glass-formers with
  Lennard-Jones like interactions. We see that
  the power law dependence holds in all cases, but the exponent
  systematically depends on the model details.}
  To emphasize the wide range over which the fractional exponent varies, we also
show the scaled plot of diffusion coefficients in Fig. \ref{all}. The
exponent values and other relevant parameters for different systems 
are collected in Table \ref{tab:table1}. We also show the temperature $T_{SEB}$
at which the Stokes-Einstein relation breaks down for $3DKA$ and $3DSqW$ model glass formers. This highlights that the power law
dependence remains valid for a wide dynamical range spanning both the
normal SE regime as well as in the supercooled, SEB regime. In this
context, we mention that the mode coupling theory (MCT) predicts a
power law dependence of the form $D \propto (T-T_c)^{-\gamma}$ for the
diffusion coefficient, in a single-component system, where $T_c$ is
the MCT transition temperature. One may reasonably expect that the
same functional form is valid for each component in a multi-component
system, thus \emph{close} to $T_{c}$. Then different components may be
related to each other {\it via} a power law \cite{gotze2008complex,kob-binder} as recently also verified \cite{spdas2018}. However, such an argument can
be applicable (i) only close to $T_c$, and (ii) provided that  $T_c$ is
the same for all components. However, even if the assumption (ii) is
valid - which is not obvious \emph{a-priori} and in fact counter
example exists \cite{murarka2003diffusion} - the MCT argument still
can not explain the power law dependence far away from $T_c$. In a
similar vein, entropy-dependent activation forms for relaxation and scaling relations between 
diffusion coefficients and appropriate entropy such as the
Rosenfeld, the Dzugutov and the Adam-Gibbs relations can not explain
the behaviour. However, the observed behaviour is a necessity for these relations to describe the individual component diffusion coefficients consistently. The Rosenfeld and the Dzugutov relations are
valid in the normal, high temperature liquid, but break down in the
supercooled, low temperature regime. The Adam-Gibbs relation, which motivated
the present survey, can not be reliably tested at high temperatures
owing to the technical difficulty in measuring configuration entropy. Thus
we highlight that the observed power law dependence between diffusion
coefficients among different components of a multi-component mixture,
should have a wider range of validity that goes beyond the theories of
glass transition. Interestingly, however, the seemingly elementary question concerning diffusivity ratios in multi-component liquids has not received theoretical attention despite a vast body of literature studying transport phenomena in varied related contexts. We hope that the present survey will stimulate interest to consider this question more clearly. 


\section{Acknowledgments}
We thank Dr. Smarjit Karmakar, Indrajit Tah, Rajshekher Das for providing 
simulation data of 2DKA, 2DMKA and 2DR10 glass formers, Prof. J\"{u}rgen Horbach 
for providing diffusion coefficient data from simulations of BKS and CHIK models and  G. Foffi for providing diffusion coefficient data 3DHS model glass former.

\bibliography{mybib}{}

\end{document}